\documentclass[onecolumn]{aastex631}

\usepackage{amsmath}
\usepackage{hyperref}
\usepackage{algpseudocode}
\usepackage{threeparttable}

\definecolor{darkbrown}{HTML}{8c4600}
\definecolor{darkblue}{HTML}{1833a1}
\newcommand{\K}{{\rm K}}
\newcommand{\m}{{\rm m}}
\newcommand{\g}{{\rm g}}
\newcommand{\s}{{\rm s}}

\submitjournal{ApJ}

\shorttitle{}
\shortauthors{Liu, Wang, Yuan, et al.}

\begin{document}


\title{Improving Radio Source Count Estimation Using Kernel Density Estimation}

\author{Luozhenhan Liu}
\affiliation{Department of Physics, School of Physics and Electronics, Hunan Normal University, Changsha 410081, China}
\affiliation{Key Laboratory of Low Dimensional Quantum Structures and Quantum Control,Hunan Normal University,Changsha 410081, China}
\affiliation{Hunan Research Center of the Basic Discipline for Quantum Effects and Quantum Technologies, Hunan Normal University, Changsha 410081, China}

\author[0000-0001-6861-0022]{Zunli Yuan}
\affiliation{Department of Physics, School of Physics and Electronics, Hunan Normal University, Changsha 410081, China}
\affiliation{Key Laboratory of Low Dimensional Quantum Structures and Quantum Control,Hunan Normal University,Changsha 410081, China}
\affiliation{Hunan Research Center of the Basic Discipline for Quantum Effects and Quantum Technologies, Hunan Normal University, Changsha 410081, China}

\author[0009-0005-1617-2442]{Wenjie Wang}
\affiliation{Department of Physics, School of Physics and Electronics, Hunan Normal University, Changsha 410081, China}
\affiliation{Key Laboratory of Low Dimensional Quantum Structures and Quantum Control,Hunan Normal University,Changsha 410081, China}
\affiliation{Hunan Research Center of the Basic Discipline for Quantum Effects and Quantum Technologies, Hunan Normal University, Changsha 410081, China}

\author{Chuanqi Li}
\affiliation{Department of Physics, School of Physics and Electronics, Hunan Normal University, Changsha 410081, China}
\affiliation{Key Laboratory of Low Dimensional Quantum Structures and Quantum Control,Hunan Normal University,Changsha 410081, China}
\affiliation{Hunan Research Center of the Basic Discipline for Quantum Effects and Quantum Technologies, Hunan Normal University, Changsha 410081, China}


\correspondingauthor{Zunli Yuan}
\email{yzl@hunnu.edu.cn}



\begin{abstract}
Radio source counts provide a fundamental census of cosmic radio emission, yet their estimation is usually based on coarse histograms that suffer from bin-choice bias, boundary effects, and survey incompleteness. We apply and rigorously evaluate kernel density estimation (KDE) as a anonparametric alternative to the conventional binned method for estimating differential radio source counts. Using simulated flux-limited samples derived from an input luminosity function model, we compare the performance of standard KDE, adaptive KDE, and traditional binning methods. Our results show that KDE-based approaches yield more accurate and stable estimates, particularly in the high-flux regime where data are sparse and conventional methods struggle. We also apply the adaptive KDE method to real observational data from the LOFAR Two-Metre Sky Survey Deep Fields. Our analysis robustly confirms the pronounced ``drop and bump" feature at sub-mJy flux densities, but also reveals that a secondary, modest bump seen in the binned data at ~ $\sim 10$ mJy is likely a binning artifact. We also demonstrate the flexibility of KDE in addressing observational incompleteness through weighted estimation, which applies weights continuously at the level of individual sources rather than averaging them in discrete bins. These strengths make KDE a powerful tool for source-count analyses in current and future radio surveys and, more broadly, in analogous studies at other wavelengths. All computations in this study are implemented with \texttt{AstroKDE}, a Python package we have developed for astronomical applications.

\end{abstract}


%
%
\section{Introduction}
\label{sec:intro}
Radio source counts quantify the number of detected extragalactic radio sources as a function of flux density, typically represented as the cumulative distribution $N(>S)$ and visualized on a log–log scale—commonly known as the ``log $N$–log $S$'' diagram. This distribution has long served as a fundamental diagnostic in extragalactic radio astronomy \citep{1966MNRAS.133..421L, 1974ApJ...188..279C}. Historically, early source count measurements were used as cosmological tests, for example, to challenge the steady-state model. A uniform Euclidean universe predicts a power-law slope of $-1.5$ in the log $N$–log $S$ relation \citep{1957PCPS...53..764S, 1967Natur.216.1076D}.

However, modern deep radio surveys have revealed a more complex behavior in the source count curve \citep{2009ApJ...690..610S, 2014MNRAS.440.2791V}. At high-flux densities, the counts are dominated by powerful radio-loud active galactic nuclei (AGNs), such as radio galaxies and quasars \citep{2001MNRAS.322..536W, 2016ApJ...831..168K}. Toward lower flux levels, the slope of the counts flattens relative to the Euclidean expectation, and an upturn is observed around the sub-mJy regime, indicating the emergence of a new population of sources \citep{2017A&A...602A...6S, 2015MNRAS.452.1263P}. In particular, star-forming galaxies (SFGs) and radio-quiet AGNs become increasingly significant contributors at flux densities around $\sim 1\,\mathrm{mJy}$ \citep{2007MNRAS.375..931M, 2012ApJS..203...15B}.

As such, the source count curve encapsulates the evolving mix of galaxy populations over cosmic time, transitioning from AGN-dominated radio emission at the bright end to emission from galaxies with intense star-formation activity at the faint end. Accurate radio source counts offer crucial insights into galaxy evolution and are instrumental in testing models of radio luminosity functions and population histories \citep{Mart_n_Navarro_2018, 2009ASPC..408..116S, 2024A&A...683A.174W}. Comparing the observed counts with predictions from evolutionary simulations (e.g., for AGN and SFG demographics) has become a standard method for constraining how these source classes evolve over time \citep{2010MNRAS.404..532M,2012ApJ...758...23C}.

Over the past two decades, large-scale surveys across multiple radio frequency bands have yielded an extensive dataset of radio sources, laying a crucial foundation for investigating their formation and cosmic evolution \citep{2012ApJ...758...23C,2017A&A...602A...6S, 2015aska.confE..67P}. Nonetheless, the accurate estimation of radio source counts remains nontrivial, as it is hindered by observational selection effects, detection thresholds, and systematic biases arising from instrumental resolution and sensitivity limitations \citep{ 2004MNRAS.355..485B,2017A&A...602A...1S, Franzen_2016}.

While the aforementioned challenges, stemming from instrumental limitations, are largely intrinsic to the hardware and must be progressively mitigated through continued advances in telescope design and data acquisition, another critical yet more tractable source of uncertainty in radio source count estimation lies in the choice of statistical methodology. Traditionally, radio source counts are estimated by binning the survey data into discrete flux density intervals and constructing a histogram (or histogram-based estimator) of the number of sources in each bin. This straightforward approach yields the familiar differential and cumulative counts used in many studies \citep[e.g.,][]{2005AJ....130.1373H, 2006MNRAS.371..963B, 2008MNRAS.383...75G}. The main drawback of binned methods is the seemingly arbitrary choice of the bin center and width, which can dramatically affect the shape of the source counts. This issue is particularly pronounced in bins that contain only a few sources \citep[e.g.,][]{1986desd.book.....S, 1998A&AS..127..335F, 2015mdet.book.....S}. As a result, the shape of the parametric form that is fit to the binned points is significantly affected \citep[e.g.,][]{2025MNRAS.539.3058G}. These drawbacks make it challenging to discern subtle astrophysical features in the source count distribution using histograms alone. Important details — such as slight inflections, plateaus, or secondary peaks in the counts — might be missed or mischaracterized if one’s binning is not optimal.

In this paper, we apply and rigorously evaluate kernel density estimation (KDE) as a robust, non-parametric alternative to the conventional binning approach. In the statistical literature, KDE is a well-established technique \citep[e.g.,][]{1986desd.book.....S, 2015mdet.book.....S, gramacki2018}, and modern statistical practice often favors kernel smoothing methods over histograms for accurate density estimation \citep[e.g.,][]{gramacki2018}. In recent years, KDE has seen increasing use in astronomy for various statistical tasks \citep{2011A&A...531A.114F, 2016MNRAS.459.2618H, 2020ApJS..248....1Y, 2022ApJS..260...10Y}, although it has not yet been widely applied to radio source count analysis. By bringing this technique into the realm of radio source counts, we aim to leverage its strengths, reducing bias associated with arbitrary binning and improving the ability to capture the true underlying distribution, to gain deeper insight into the radio source counts.

Throughout the paper, we adopt the Lambda cold dark matter (\(\Lambda\)CDM) cosmology with the parameters $\Omega_m = 0.3$, $\Omega_\Lambda = 0.7$, and $H_0 = 70 \, \text{km s}^{-1} \, \text{Mpc}^{-1}$.

%
%

\section{Methods}
\label{sec:Kernel density estimation}
\subsection{KDE}
\label{sec:intro:structure}
Suppose we observe \(n\) points in a one-dimensional (1D) space, \(X = \{X_1, X_2, \dots, X_n\}\). Assume that \(X\) comes from an unknown probability density function \(f(x)\). The goal of KDE is to estimate \(f\). Given \(X\), the classical fixed-bandwidth kernel estimate of \(f\) is written as
\begin{equation}
	\hat{f}(x) = \frac{1}{nh}\sum_{i=1}^n K\left(\frac{x - X_i}{h}\right),
	\label{eq:kde1}
\end{equation}
\noindent
where \(K\) is the kernel function, \(h\) represents the bandwidth, and \(n\) is the sample size.

It is widely accepted that the choice of bandwidth is more important than the choice of kernel function \citep{ mokkadem2006largemoderatedeviationsprinciples, biroli2024kerneldensityestimatorslarge, gramacki2018}. Throughout this study, we employ the Gaussian kernel
\begin{equation}
\
K(x) = \frac{1}{\sqrt{2\pi}} \exp\left(-\frac{x^2}{2}\right),
\
\end{equation}
which is widely used in most KDE applications.

\subsection{Bandwidth Selection}
\label{sec:intro:structure}
In KDE, the bandwidth parameter plays a central role in controlling the smoothness of the estimated density, as it determines the spatial extent of influence for each observation.
Following \citet{2020ApJS..248....1Y}, we employ the likelihood cross-validation (LCV) criterion to find optimal bandwidths. The LCV thinks about the estimated density itself as a likelihood function, and the LCV objective function is given by
\begin{equation}
	\mathrm{LCV}(h) = \frac{1}{n} \sum_{i=1}^n \log \hat{f}_{-i}(x_i, h),
\end{equation}
where $\hat{f}_{-i}(x_i; h)$ is the leave-one-out estimator:
\begin{equation}
	\hat{f}_{-i}(x_i, h) = \frac{1}{(n-1)h} \sum_{\substack{j=1 \\ j \ne i}}^n K\left( \frac{x_i - x_j}{h} \right).
\end{equation}
For a sample of $n$ observed data points $\{x_i\}_{i=1}^n$, the $\hat{f}_{-i}(x_i; h)$ is the KDE at $x_i$ using all data except $x_i$, evaluated with a candidate bandwidth $h$.
The optimal bandwidth $h_{\mathrm{opt}}$ is then obtained by maximizing the cross-validation function:
\begin{equation}
	h_{\mathrm{opt}} = \arg\max \mathrm{LCV}(h).
\end{equation}


\subsection{Adaptive KDE}
\label{sec:methods:experiment}

To enhance the flexibility of KDE, an adaptive approach is often adopted, in which the bandwidth \( h \) varies with local data density \citep{Breiman1977}. Specifically, smaller bandwidths are used in regions of high data concentration, while larger bandwidths are applied in sparser areas, thereby improving estimation accuracy relative to the fixed-bandwidth case \citep{davies2018statistics, Botev_2010}. The 1D adaptive KDE estimator is given by \citep[e.g.,][]{sain2002computational}:
\begin{equation}
	\hat{f}_a(x) = \frac{1}{n} \sum_{i=1}^{n} \frac{1}{h_i} K\left( \frac{x - X_i}{h_i} \right).
\end{equation}

I.\textcite{abramson1982bandwidth} proposed that the bandwidth can be set inversely proportional to the square root of the true density. Since the true density \( f \) is generally unknown, a pilot estimate is used in practice. Following \textcite{1986desd.book.....S}, this approach can be implemented as
\begin{equation}
	h_i = \frac{h_{0}}{[ \hat{f}(X_i) ]^{\beta}},
    \label{eq:h_i}
\end{equation}

\noindent
where \( \hat{f}(X_i) \) is a pilot estimate obtained via fixed-bandwidth KDE (e.g., Equation~(\ref{eq:kde1})), and \( h_0 \) is the global smoothing parameter also known as the global bandwidth \citep{davies2018statistics}. The exponent \( \beta \) controls the degree of adaptivity and is commonly set to 0.5. In this work, however, we treat \( \beta \) as a free parameter and determine its optimal value by maximizing the likelihood-based cross-validation criterion.

\subsection{Boundary effects for KDE}
\label{sec:methods:gas}

In radio astronomy, samples used for source count estimation are typically flux limited, meaning they are left truncated at a detection threshold $S_{\min}$. For such truncated data, standard KDE cannot be applied directly, as it suffers from significant boundary bias \citep[e.g.,][]{gasser1979kernel, marron1994kernel, hall2002nonparametric, marshall2010kernel, maleca2014adaptive, davies2018statistics} near the flux limit.

More generally, consider a set of observations $\{X_i\}_{i=1}^n$ truncated from below at $x_0$. To correct for the boundary effect in KDE, one common approach is the reflection method \citep[e.g.,][]{jones1993simple}. The idea is to augment the sample by introducing a mirrored data point for each observation, defined as $X_i^* \equiv 2x_0 - X_i$. The KDE estimator for the truncated sample can then be written as
\begin{equation}
	\label{eq:kder}
    \hat{f}_{\text{ref}}(x) = \frac{1}{nh} \sum_{i=1}^{n}
    \left[
    K\!\left( \frac{x - X_i}{h} \right) +
    K\!\left( \frac{x + X_i - 2x_0}{h} \right)
    \right],
\end{equation}

where the second term, involving the reflected data points $X_i^*$, represents their contribution to the estimate near the truncation boundary. Similarly, in the case of adaptive KDE, the reflection method is also applicable:
\begin{equation}
	\label{eq:kdera}
    \hat{f}_{\text{ra}}(x) = \frac{1}{n} \sum_{i=1}^{n} \frac{1}{h_i} \left[ K\left( \frac{x - X_i}{h_i} \right) + K\left( \frac{x + X_i - 2x_0}{h_i} \right) \right],
\end{equation}
where \( h_i \) is given by Equation \ref{eq:h_i}. The reflection method allows us to correct for the underestimation of density near boundaries and produce more reliable estimates across the entire domain.

%
%
\subsection{Estimating the Source Count via KDE}
\label{sec:ESTIMATING THE SOURCE COUNT VIA KDE}

The differential form of radio source counts is expressed as
\begin{equation}
\label{dnds}
    \hat{n}(S)=\frac{\mathrm{d}N}{\Omega\,\mathrm{d}S},
\end{equation}

\noindent
 where $\Omega$ denotes the solid angle of the survey and $S$ is the flux density. Physically, this quantity represents the number of radio sources per unit flux interval per unit solid angle.

To facilitate interpretation and comparison across different surveys, the differential counts are often multiplied by $S^{2.5}$, yielding a Euclidean-normalized form \citep[e.g.,][]{ 1974ApJ...188..279C, 2011hea..book.....L}. This transformation accounts for the expected behavior in a static, nonevolving Euclidean universe, where source counts follow a power-law slope of $-1.5$ in the cumulative distribution. Deviations from this expectation provide valuable insights into cosmic evolution and the emergence of different source populations, including the transition from AGN-dominated to SFG populations at faint flux levels \citep{2017A&A...602A...1S}.

It is often more convenient to perform the analysis in logarithmic flux space, in which case the differential source counts transform as
\begin{equation}
	\frac{dN}{\Omega dS}=\frac{dN}{d\log S}\cdot\frac{d\log S}{\Omega dS}=\frac{1}{\Omega S\ln10}\cdot\frac{dN}{d\log S}.
\end{equation}
Let \( x \equiv \log S \), and \( X_i \) denote the logarithmic flux density of the \( i \)th source in the sample. The density of \( x \), \( \hat{f}(x) \), can then be estimated using KDE, for example, via Equation~(\ref{eq:kde1}). In radio astronomy, source count studies typically rely on flux-limited samples, which are truncated below a survey-dependent detection threshold \( S_{\lim} \). As a result, the most appropriate approach to estimating \( \hat{f}(x) \) in practice is to adopt a boundary-corrected KDE method, as formulated in Equations~(\ref{eq:kder}) and~(\ref{eq:kdera}), corresponding to the fixed and adaptive bandwidth cases, respectively.

Given that
\begin{equation}
\frac{\mathrm{d}N}{ \mathrm{d} x} \equiv {n} \cdot \hat{f}(x) \equiv
\frac{\mathrm{d}N}{ \mathrm{d} \log S},
\end{equation}
we obtain the KDE-based estimate of the source counts as:
\begin{equation}
\label{dndskde}
\hat{n}(S)=	\frac{1}{h\Omega S\ln10}\cdot\sum_{i=1}^{n}
\left[K\left(\frac{\log S-\log S_{i}}{h}\right)
+ K\left(\frac{\log S+\log S_{i}-2\log S_{\lim}}{h}\right)
\right].
\end{equation}
For the adaptive bandwidth case, a similar expression follows directly by replacing the fixed bandwidth \( h \) with a locally varying \( h_i \), as in Equation~(\ref{eq:kdera}).
All KDE-related computations in this study were performed using \texttt{AstroKDE}, a Python package we developed specifically for astronomical applications. This package will soon be released on the Python Package Index . Throughout this paper, we denote the estimated differential source counts as \(\hat{n}_{\mathrm{bin}}\) for the binned method, \(\hat{n}_{\mathrm{KDE}}\) for fixed-bandwidth KDE, and \(\hat{n}_{\mathrm{aKDE}}\) for adaptive KDE.


\section{results}
\subsection{Application of KDE to Simulated Samples}
\label{sec:APPLECATION OF KDE TO SIMULATED  SOURCE SAMPLES}

To evaluate the effectiveness of the KDE-based approach for estimating radio source counts, we construct a controlled simulation framework. Specifically, we generate mock flux-limited samples from an input radio luminosity function model \citep[][Model A]{2017ApJ...846...78Y}, allowing us to test and compare the performance of the KDE method against the conventional binned method.

Given that the underlying luminosity function $\Phi(z, L)$ is known by construction, we are also able to compute the “true” differential source counts through the following equation:
\begin{equation}
	\hat{n}(S) = 4\pi \frac{c}{H_0}
	\times \int_{z_1}^{z_2}
	\frac{\Phi\left( z, L( z, S)\right) D_L^4(z)\, dz}
	{(1+z)^{3-\alpha} \sqrt{\Omega_M (1+z)^3 + \Omega_\Lambda}},
		\label{eq:dnds_RLF}
\end{equation}

\noindent
where \( c \) is the speed of light, \( D_L(z) \) is the luminosity distance, \( \alpha\) is the spectral index (assumed to be a fixed value of 0.75), and \( z_1 \), and  \( z_2 \) represent the range of integration in redshift \citep[e.g.,][]{2016A&ARv..24...13P,2017ApJ...846...78Y}.

\begin{figure*}[t]
	\centering
	\includegraphics[width=\textwidth]{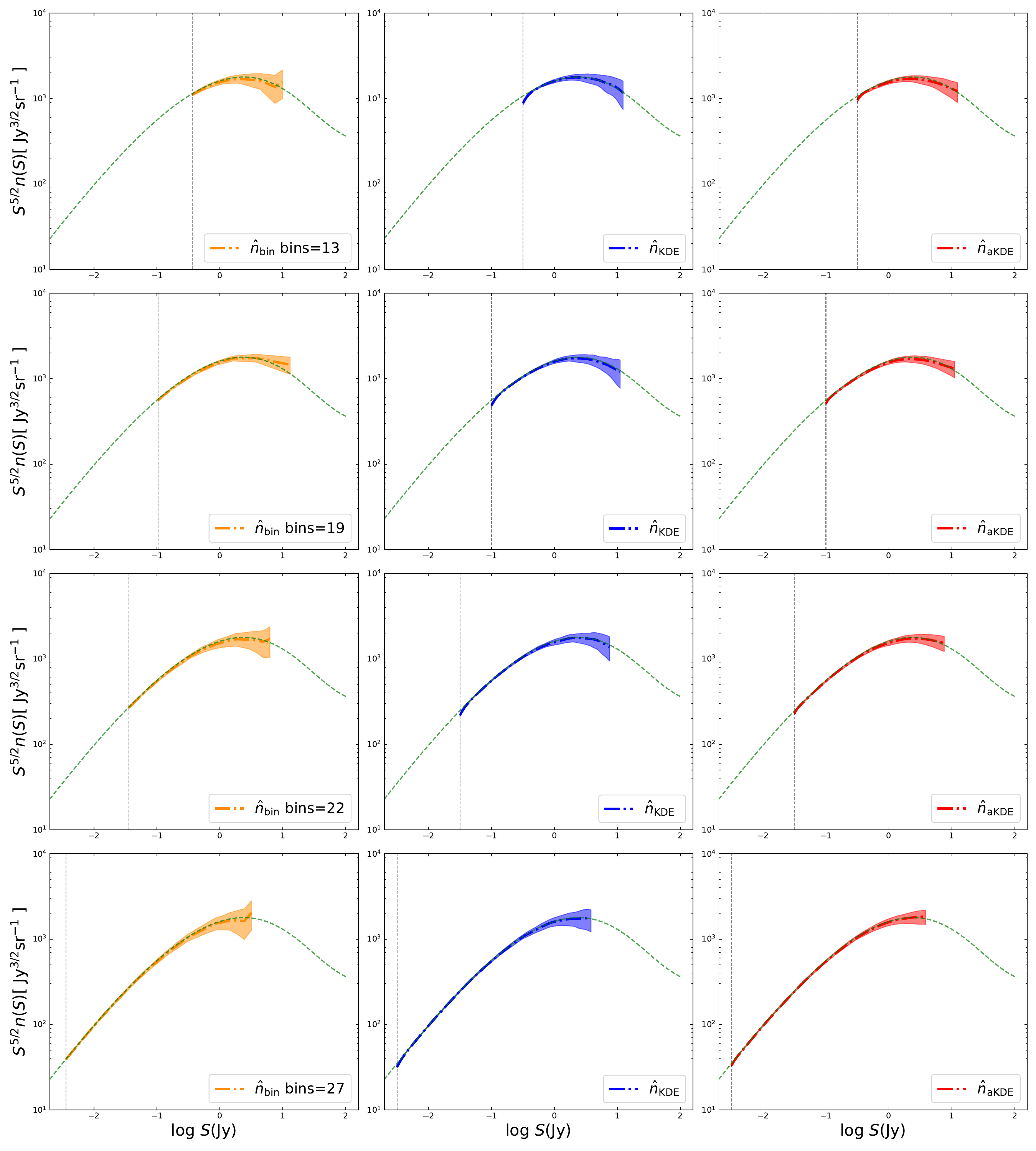}
	\caption{
Mean differential source counts derived from 200 simulated realizations, estimated using three methods: the traditional binned approach (orange), standard KDE (blue), and adaptive KDE (red). In each panel, the green dashed line shows the true source counts computed from Equation~(\ref{eq:dnds_RLF}), and the shaded regions indicate the 1$\sigma$ dispersion across the 200 samples. The flux limit adopted in each case is marked by a vertical dotted line.
In the binned method, we divide the logarithmic flux density space into equal-width bins according to Scott’s rule, with the number of bins indicated in each panel.}
	
	\label{fig:200timesscerr}
\end{figure*}

In the simulation analysis, we consider four different flux limits, as summarized in Table~\ref{tab:sample}. For each flux limit, we generate 200 independent mock samples, all with identical sample sizes at a given limit. Each set of simulated samples is then used to estimate the source counts using three different methods: KDE, adaptive KDE, and the binned approach.
In the binned method, we divide the logarithmic flux space into equal-width bins, with the bin width determined by Scott’s rule \citep{Scott1979}:
\begin{equation}
    h = 3.5 \cdot \frac{\sigma_{\log S}}{n^{1/3}},
\end{equation}
where \( h \) is the bin width, \( \sigma_{\log S} \) is the standard deviation of the sample in logarithmic flux space, and \( n \) is the number of sources in the sample.
Each set of simulated samples is then used to estimate the source counts via three different approaches: KDE, adaptive KDE with locally varying bandwidths, and the conventional binned method. In Figure~\ref{fig:200timesscerr}, we present, for each flux limit, the mean differential source counts derived from the 200 simulated samples using the three different estimation methods. The shaded regions indicate the corresponding 1$\sigma$ intervals, representing the sample-to-sample variation within each method. In each panel, the ``true" differential source counts, computed from Equation~(\ref{eq:dnds_RLF}), are shown as a green dashed–dotted line.

\begin{table}[htbp]
	\centering
	\caption{Parameters of the simulated source samples.}
	\label{tab:sample}
	\begin{tabular}{lcccc}
		\hline
		\hline
		Flux Limit $S_{\mathrm{lim}}$ (Jy) & $10^{-0.5}$ & $10^{-1.0}$ & $10^{-1.5}$ & $10^{-2.5}$ \\
		\hline
		Survey area $\Omega$ (sr) & 0.4134 & 0.3731 & 0.1995 & 0.0693 \\
		Number of sources $n$ & 20,007 & 10,013 & 6403 & 2014 \\
		\hline
	\end{tabular}
\end{table}

\begin{figure*}[t]
	\centering
	\includegraphics[width=0.75\textwidth]{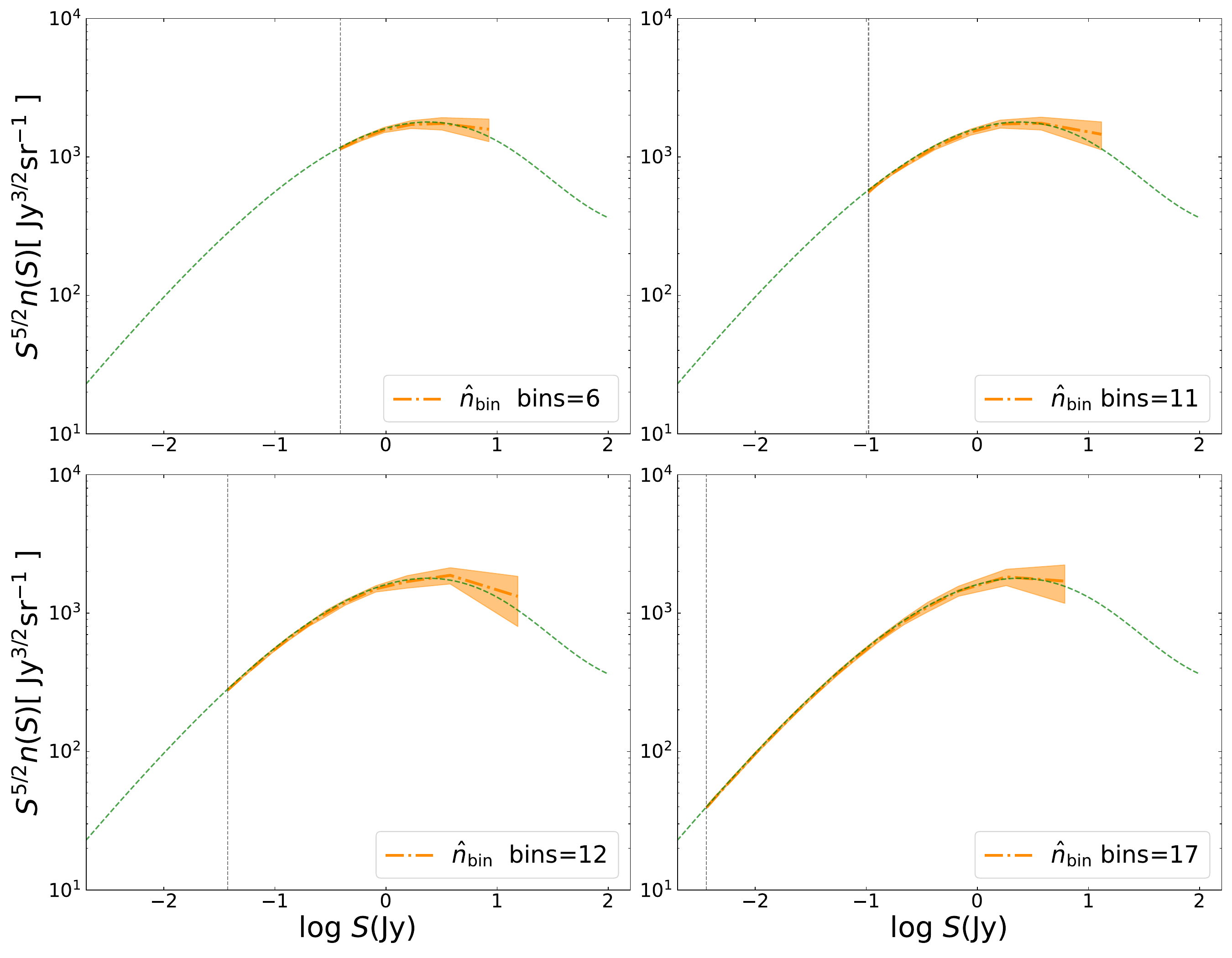}
	\caption{
Similar to Figure~\ref{fig:200timesscerr}, except that here the differential source counts are estimated using the adaptive binning method based on the Bayesian Blocks algorithm \citep{Scargle2013}. In this approach, the bin widths vary with the local data density.
    }		
	\label{fig:scbybayesian}
\end{figure*}

Figure~\ref{fig:200timesscerr} provides a direct comparison of the performance of the three estimation methods across four flux limits. For each case, the KDE-based approaches—both with fixed and adaptive bandwidths—yield mean differential source counts that closely follow the input source counts, with excellent agreement over the entire flux range. The associated $1\sigma$ dispersions, shown as shaded regions, are consistently narrow, demonstrating the statistical stability of the KDE methods.

While both KDE variants perform well, the adaptive KDE shows slightly better behavior near the low-flux boundary and in the bright-source regime, where the number of detections is low. In these regions, it tends to yield marginally smaller uncertainties due to its ability to locally adjust smoothing based on data density.

Compared to the KDE-based estimators, the binned method exhibits somewhat larger $1\sigma$ dispersions across flux limits, reflecting its greater sensitivity to sample variance. At the bright end—where the number of detections is small—the average counts from the binned method tend to deviate from the true values. By contrast, the KDE approaches yield average counts that closely track the true distribution, with the estimated curves nearly overlapping the reference counts. This demonstrates the ability of KDE to provide more stable and unbiased estimates, particularly in regimes where conventional binning becomes unreliable.

It is well known that the results of binned estimators are sensitive to the choice of binning scheme. To examine this dependence, we also adopt an alternative binning strategy based on the Bayesian Blocks algorithm \citep{Scargle2013}, which adaptively adjusts the bin widths according to the local data density. The resulting source count estimates are presented in Figure~\ref{fig:scbybayesian}, illustrating the performance of this variable-bin approach. Compared with the fixed-width scheme, it shows moderate improvement by adapting to the data distribution; however, the estimated counts can still deviate from the true values, indicating that binning-related biases are not fully eliminated. The advantage of the variable-bin method becomes more apparent for large samples, where the algorithm has sufficient statistical support to adjust bin widths flexibly and to capture genuine variations in the source counts.

To quantitatively evaluate the performance of KDE in recovering the underlying radio source counts, we define a metric \( d_n \) that measures the average fractional deviation in logarithmic space between the estimated counts and the theoretical reference. Specifically,
\begin{equation}
	d_n = \frac{1}{n} \sum_{i=1}^{n} \left| \frac{  \log_{10} \hat{n}(S_i)-\log_{10} n_{\mathrm{true}}(S_i)}{\log_{10} n_{\mathrm{true}}(S_i)} \right|,
  \label{dF}
\end{equation}

\noindent
where \( \hat{n}(S_i) \) denotes the differential source counts estimated from the simulated samples at flux density \( S_i \), and \( n_{\mathrm{true}}(S_i) \) is the corresponding theoretical value computed from the input luminosity function via Equation~(\ref{eq:dnds_RLF}).

For the binned method, we treat the estimate as a piecewise constant function \citep{2022ApJS..260...10Y}. Each source $s_i$ is assigned the source count value at the center of the bin it falls into:
$$
\hat{n}(S_i) = \hat{n}_{\mathrm{bin}}(\bar{S}_j),
$$
where $\bar{S}_j$ is the central flux of the bin containing $S_i$. This allows the discrepancy metric $d_n$ of Equation (\ref{dF}) to be applied uniformly across all methods.

With the discrepancy metric $d_n$ defined above, we proceed to quantitatively evaluate the three estimation methods. For each flux limit, we compute $d_n$ for all 200 simulated realizations using the binned method, standard KDE, and adaptive KDE.

\begin{figure*}[htbp]
	\centering
	\includegraphics[width=0.9\textwidth]{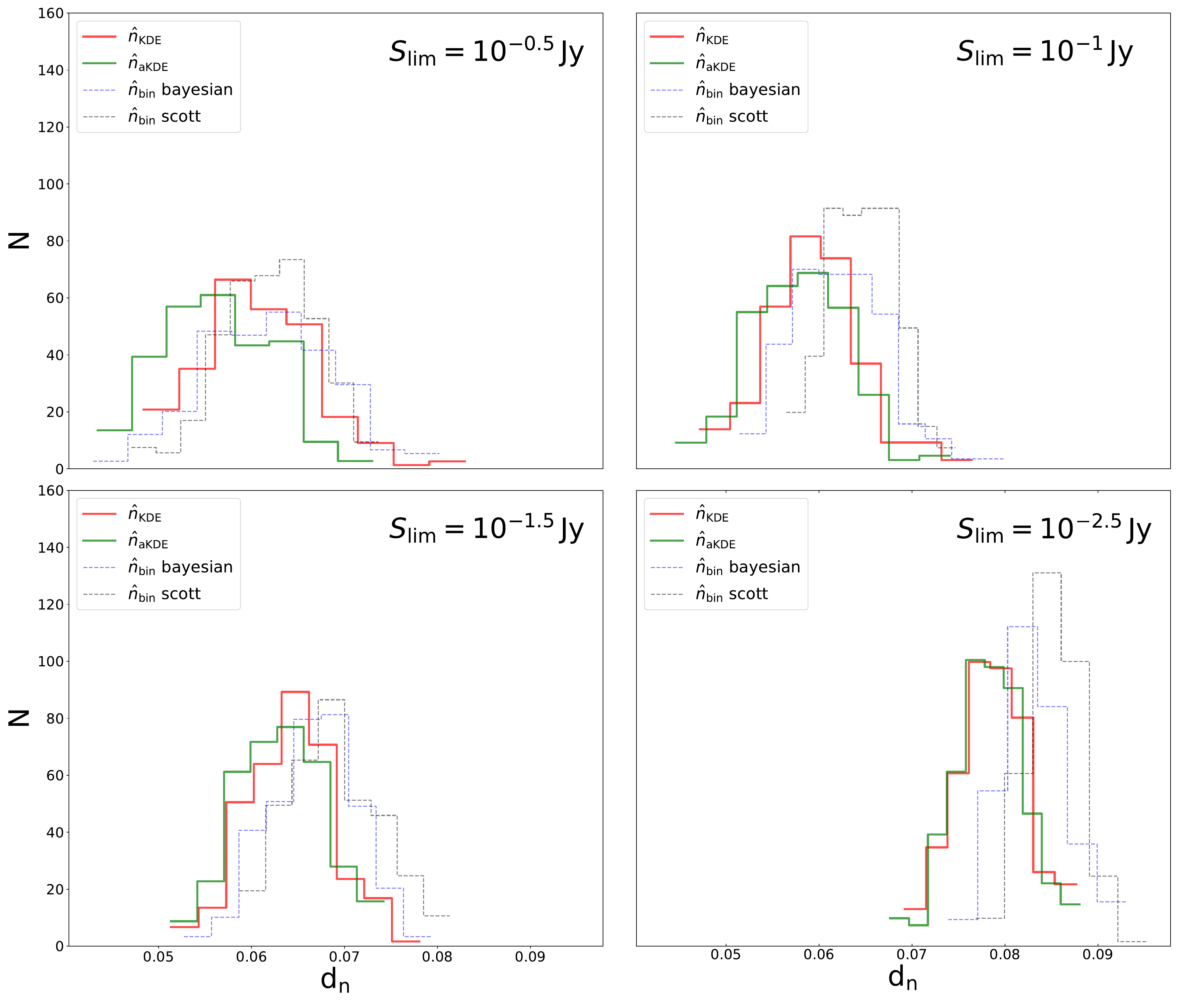}	
\caption{Distributions of the discrepancy metric \( d_n\) for different source count estimators, based on 200 simulated samples at each flux limit. Each panel corresponds to a different flux threshold.}	
	\label{fig:relativeerr}
    \end{figure*}

Figure~\ref{fig:relativeerr} illustrates the distributions of the $d_n$ values obtained from 200 simulations under each flux limit, offering insight into the stability and spread of each estimator. Complementarily, Table~\ref{tab:ans} summarizes the central tendency of these distributions by reporting the mean $d_n$ for each method, providing a compact comparison of their overall accuracy. As shown in Figure~\ref{fig:relativeerr} and Table~\ref{tab:ans}, both KDE-based approaches outperform the binned method across all flux limits, yielding smaller $d_n$ values.

\begin{table}[htbp]
    \centering
    \caption{Mean \( d_n\).}
    \label{tab:ans}
    \begin{threeparttable}
        \begin{tabular}{lcccc}
            \hline\hline
            Method & \(10^{-0.5}\) Jy & \(10^{-1.0}\) Jy & \(10^{-1.5}\) Jy & \(10^{-2.5}\) Jy \\
            \hline
            Binned Scott   & 0.0619 & 0.0646 & 0.0689 & 0.0852 \\
            Binned Bayesian  & 0.0618 & 0.0622 & 0.0666 & 0.0834 \\
            KDE       & 0.0592 & 0.0593 & 0.0642 & 0.0786 \\
            Adaptive KDE & 0.0554 & 0.0576 & 0.0631 & 0.0783 \\
            \hline
        \end{tabular}
    \end{threeparttable}
\end{table}

\subsection{Application of KDE to Real Radio Survey Data}

In this section, we apply the KDE-based estimators to real observational data from the LOFAR Two-Metre Sky Survey(LoTSS) Deep Fields presented by \citet{2021A&A...648A...5M}. The sample consists of radio sources detected at 150\,MHz in the Lockman Hole, Boötes, and ELAIS-N1 fields, covering effective survey areas of 10.3, 8.6, and 6.7\,deg$^{2}$, respectively. Following the procedures described by \citet{2021A&A...648A...5M}, the raw catalogs were quality-checked to remove spurious detections and to deblend confused sources, resulting in cleaned source lists containing 31,163, 19,179, and 31,645 objects in the three fields, respectively.

To correct for incompleteness, we adopt the completeness estimates derived by \citet{2023MNRAS.523.6082C}, who quantified the detection probability as a function of flux density by injecting simulated sources into the LoTSS Deep Fields images and reextracting them using the same source-finding pipeline. These completeness functions were applied as statistical weights to individual sources in our analysis, ensuring that our estimates consistently account for selection effects and remain directly comparable to the fixed-bin results reported by \citet{2021A&A...648A...5M}.

\begin{figure}[htbp]  
    \centering
    \includegraphics[width=1.0\textwidth]{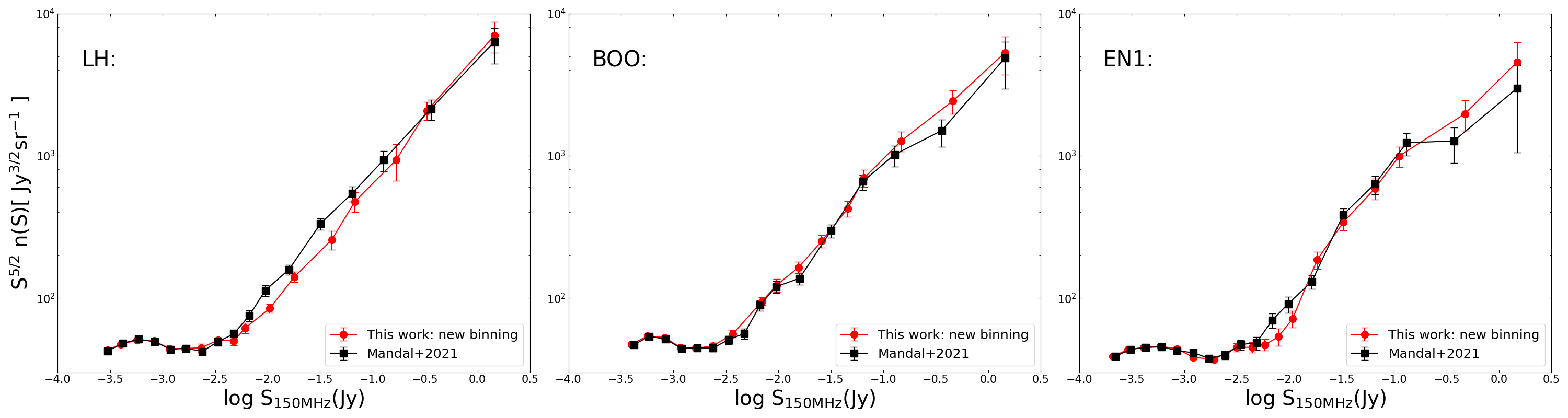} 
    \caption{
Differential source counts for the three LoTSS Deep Fields (Lockman Hole, Boötes, and ELAIS-N1) obtained with the traditional binned estimator. Black filled squares show the results of \citet{2021A&A...648A...5M}, and red filled circles show our own estimates. The binning scheme follows \citet{2021A&A...648A...5M} with slight shifts in several bin edges, illustrating the sensitivity of histogram-based estimators to bin placement.}
    \label{fig:bin} 
\end{figure}

\begin{figure}[htbp]  
    \centering
    \includegraphics[width=1.0\textwidth]{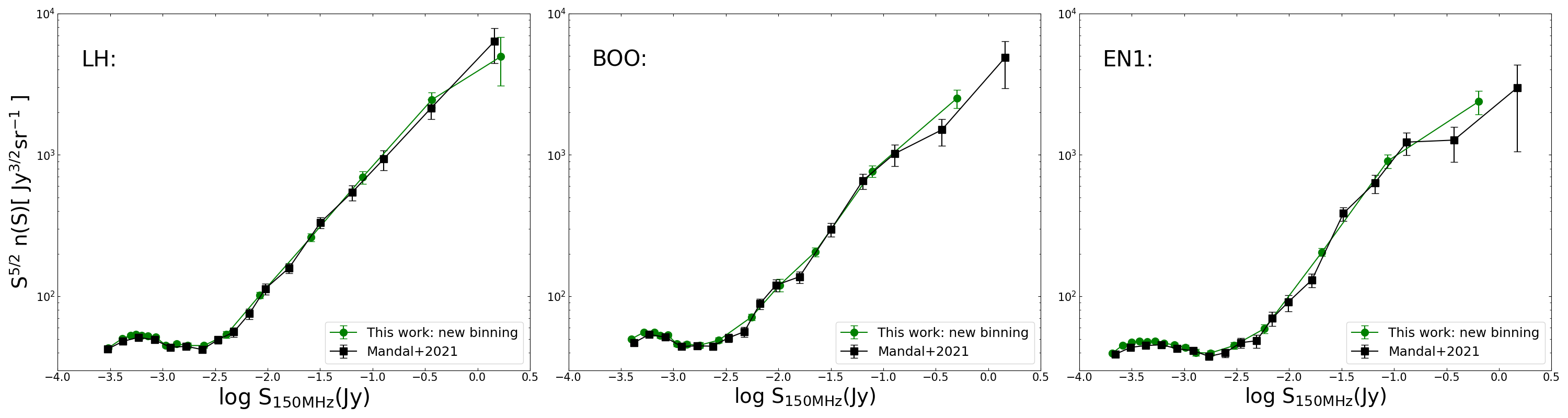} 
\caption{
Similar to Figure~\ref{fig:bin}, black squares denote the results of \citet{2021A&A...648A...5M}, and dark-green circles show our estimates using the Bayesian Blocks adaptive binning algorithm.}

    \label{fig:bayesian} 
\end{figure}

\begin{figure}[htbp]  
    \centering
    \includegraphics[width=1.0\textwidth]{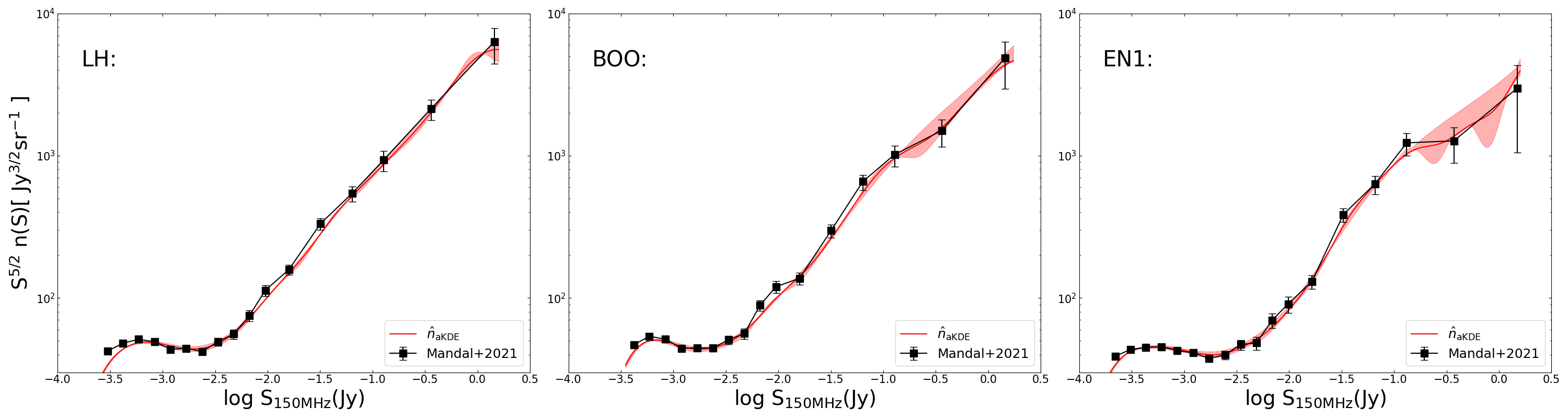} 
\caption{
    Differential source counts for the three LoTSS Deep Fields, estimated using the adaptive KDE method (solid red lines). The pale-red shaded region indicates the $3\sigma$ uncertainty band for the KDE estimate. For comparison, the binned results from \citet{2021A&A...648A...5M} are shown as black squares.}

    \label{fig:akde} 
\end{figure}

Figure~\ref{fig:bin} shows the differential source counts derived using the traditional binned estimator for the three LoTSS Deep Fields: the Lockman Hole, Boötes, and ELAIS-N1. In each panel, the black filled squares represent the results reported by \citet{2021A&A...648A...5M}, while the red filled circles show our own binned estimates. The binning scheme adopted here follows that of \citet{2021A&A...648A...5M}, but with slight adjustments to the starting positions of several bins in order to illustrate the sensitivity of the results to the binning configuration. As can be seen, even these modest changes in bin boundaries lead to noticeable differences in the estimated source counts, highlighting the inherent instability of histogram-based estimators with respect to bin placement.

The instability demonstrated in Figure~\ref{fig:bin} highlights the risks of empirical or ad hoc bin selection, as used by \citet{2021A&A...648A...5M}. We therefore test a more advanced adaptive binning strategy based on the Bayesian Blocks algorithm \citep{Scargle2013}. This test is to determine if an objective, data-driven binning method can resolve the instabilities and confirm the features seen in the \citet{2021A&A...648A...5M} data. As shown in Figure~\ref{fig:bayesian}, this adaptive method yields source counts that are in closer overall agreement with the results of \citet{2021A&A...648A...5M} (black filled squares) than our perturbed bins. Importantly, our Bayesian Blocks estimate clearly confirms the pronounced ``drop and bump" feature at sub-mJy flux densities (around $S \sim 2$ mJy, or $\log S \approx -2.7$) which \citet{2021A&A...648A...5M} reported as their main finding. However, some discrepancies remain in other, finer details. In particular, for the Boötes field, the \citet{2021A&A...648A...5M} binned data also display a separate, more modest bump around $\log S \approx -2$ (i.e., $\sim 10$ mJy), a feature not explicitly discussed in their paper. Our new Bayesian Blocks estimate does not reproduce this specific, modest bump at $\log S \approx -2$.

\begin{figure}[htbp]  
    \centering
    \includegraphics[width=0.3\textwidth]{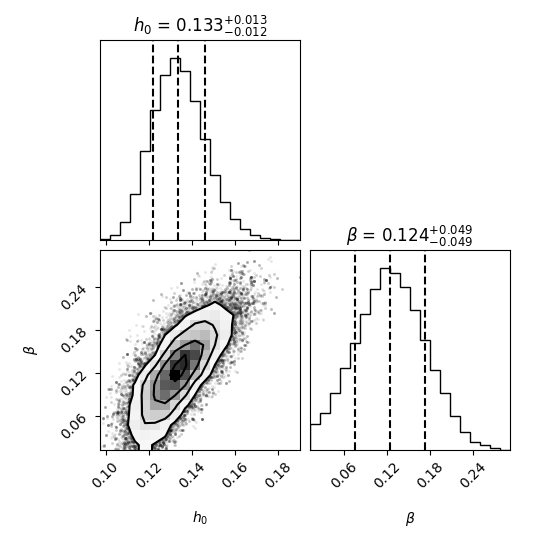} 
    \includegraphics[width=0.3\textwidth]{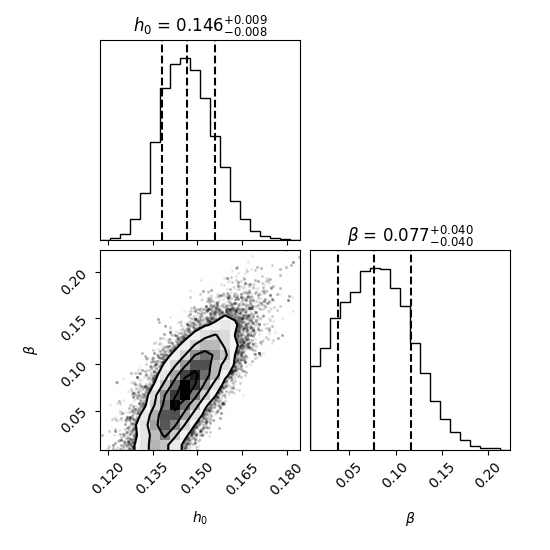}
    \includegraphics[width=0.3\textwidth]{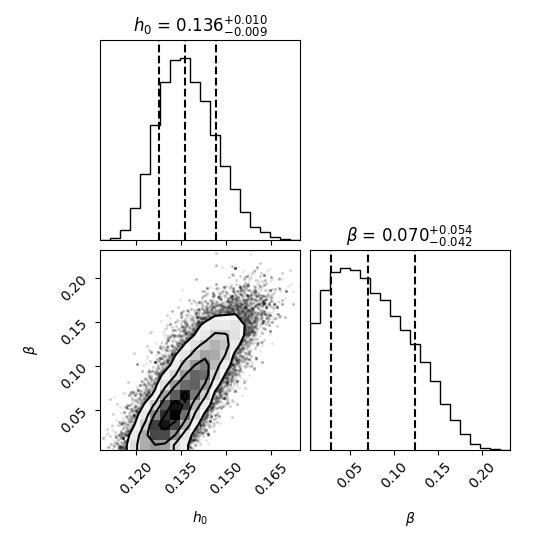}
\caption{
Posterior distributions of the adaptive KDE parameters $h_{0}$ and $\beta$, derived with the routine \texttt{AstroKDE}. Panels from left to right show the three LoTSS Deep Fields: Lockman Hole, Boötes, and ELAIS-N1.
}
    \label{fig:post_akde} 
\end{figure}

To provide an estimate free from any binning-related biases, we apply our adaptive KDE estimator. Figure~\ref{fig:akde} presents the results. The continuous KDE estimates are in good agreement with the binned results on a global scale,
and they also robustly recover the main ``drop and bump" feature at sub-mJy levels.
However, in the Boötes field, the KDE estimate (red line) is smooth and, consistent with the independent Bayesian Blocks result, also does not show the modest secondary bump seen in the \citet[data, black squares]{2021A&A...648A...5M}  around $\log S \approx -2$. This finding, combined with the independent Bayesian Blocks result illustrates that an apparent `bump'  feature in binned data can sometimes be an artifact of a specific binning choice, rather than a genuine feature of the underlying source distribution. To assess the robustness of the adaptive KDE estimation, Figure~\ref{fig:post_akde} displays the posterior distributions of the bandwidth parameter $h_{0}$ and the adaptivity parameter $\beta$ for the three LoTSS Deep Fields, showing consistent and well-constrained solutions across all fields.

\section{Discussion}
\label{sec:Discussion}
The results of our analysis demonstrate that KDE offers a flexible and statistically robust alternative to the traditional binned approach for estimating radio source counts. This advantage becomes increasingly important as radio astronomy enters ``the era of precision cosmology." Modern deep radio surveys are producing unprecedented volumes of high-sensitivity data, and the demands on statistical methodologies are correspondingly growing.

Recent observing programs, such as  LoTSS (\citealt{2019A&A...622A...1S, 2022A&A...659A...1S}),Very Large Array (VLA)-COSMOS \citep{2017A&A...602A...1S}, EMU with ASKAP \citep{2020AAS...23632206K}, and MeerKAT’s MMIGHTEE \citep{2016mks..confE...6J}, have already demonstrated the potential of deep surveys to reveal faint and complex radio populations. The upcoming Square Kilometre Array (SKA) and its precursors are expected to detect millions of sources, including rare and morphologically diverse classes of radio galaxies. In this
context, the ability to extract accurate, high-resolution source
count distributions—without the bias introduced by arbitrary
binning—is essential.

KDE-based methods are particularly well suited to this task. By providing continuous density estimates and avoiding coarse binning, KDE enables the detection of fine-scale variations in the source count distribution. A striking example is the sub-mJy bump in the 150 MHz counts reported by \citet{2021A&A...648A...5M}, a feature that had eluded earlier analyses based on shallower data or conventional histograms. Identifying such patterns is crucial for understanding the emergence of new source populations and tracing the transition from AGN-dominated to star formation–dominated regimes.

KDE-based methods are particularly well suited to this task. By providing continuous density estimates and avoiding coarse binning, KDE enables a more robust detection of fine-scale variations in the source count distribution. Conventional histograms, in contrast, are highly sensitive to bin placement, which can not only smooth over real physical features but also introduce artificial bumps or dips that are not present in the underlying data (as we demonstrate in Section 3.2). Identifying such patterns robustly is crucial for understanding the emergence of new source populations and tracing the transition from AGN-dominated to star-formation-dominated regimes.

\subsection{KDE for Large Surveys} \label{sec:big_data}
It is worth noting, however, that as the sample size increases, the performance of binned estimators gradually stabilizes, since larger numbers of sources per bin reduce the effects of shot noise and make the results less sensitive to the specific choice of binning. This naturally raises the question of whether KDE is still necessary in the high-data regime anticipated for SKA and its pathfinders. We argue that the answer is yes. As astronomy enters the era of precision cosmology, the dominant challenge is shifting from mitigating statistical errors (which large $N$ solves) to eliminating systematic biases (which large $N$ does not solve). KDE’s advantages are critical in this new regime for several reasons.

First, and most importantly, a large total $N$ does not eliminate local sparsity. Even in massive surveys, the high-flux tail of the distribution will always be sparsely populated. The superior accuracy of KDE in this regime is essential for a robust understanding of the rare and powerful sources that often hold key physical insights.

Second, KDE provides a high-fidelity, continuous estimate that avoids the systematic biases inherent in discrete binning. These artifacts—such as sensitivity to bin width and center, or boundary biases (as discussed in Section 2.4)—are not eliminated by large statistics. A continuous KDE estimate is therefore fundamentally superior for detecting the very subtle astrophysical features (e.g., inflections or small-scale deviations in the source count curve) that precision data will enable us to find, but which a discrete histogram can easily "bin over" or mischaracterize.

\subsection{Incompleteness Correction} \label{sec:incompleteness}
In addition, a notable strength of KDE lies in its flexibility to
accommodate observational incompleteness—a common challenge in deep radio surveys. Recent large-area observations such as the VLA 3~GHz survey have shown that completeness can vary significantly with flux density and sky position \citep[e.g.,][]{2017A&A...602A...6S}. In such cases, KDE provides a natural framework for correction through weighted density estimation.  While both KDE-based and traditional binned estimators apply the same completeness weights to individual sources, the key difference lies in the discretization step inherent to binning. In binned estimators, the individually weighted sources are summed into discrete, predefined bins. This process, while correctly summing the weights, is still subject to all the previously discussed binning biases (e.g., sensitivity to bin boundaries), which can be particularly problematic if the completeness function varies rapidly across a bin \citep[also see][]{2013Ap&SS.345..305Y}. In contrast, KDE applies the weights continuously at the level of individual data points, allowing local variations in completeness to be captured without discretization effects. By assigning each source an inverse completeness weight based on its flux (or other relevant parameters), the KDE can recover the intrinsic distribution without relying on fixed-bin edges or complex bin-wise adjustments. This weighting scheme integrates seamlessly with both fixed and adaptive KDE formulations \citep{2022ApJS..260...10Y}, preserving their nonparametric nature while enhancing robustness under realistic survey conditions.

This adaptability to varying data completeness further highlights the value of KDE in modern radio astronomy, where factors like survey depth, detection thresholds, and instrumental sensitivity often vary across the sky. Consequently, the flexibility of KDE inheterogeneous datasets is not
only beneficial for boundary correction and density resolution, but also critical for correcting sample selection biases.

Taken together, these strengths highlight the growing relevance of KDE-based methods in the analysis of data from the next generation of radio surveys. As the community moves toward data-driven modeling of source populations, nonparametric techniques like KDE will play a key role in extracting astrophysical insights from increasingly complex and voluminous datasets. We therefore advocate for the adoption of KDE as a standard tool for radio source count analysis in the SKA era and beyond, facilitated by accessible tools such as \texttt{AstroKDE}.

\subsection{Limitations and Future Work} \label{sec:limitations}
Finally, we must address an important limitation of the present study. This work, in common with many traditional source count analyses, has not explicitly accounted for the impact of flux density measurement errors. Near the faint-end flux limit, where the true source counts rise steeply, measurement uncertainties lead to the well-known Eddington bias \citep{1940MNRAS.100..354E}, which can cause a systematic overestimation of the observed counts. This observational bias is a complex problem that is distinct from the algorithmic boundary bias (an underestimation) that our KDE reflection method is designed to correct.

The scope of this paper is to provide a rigorous comparison between binned and KDE methods for estimating the measured source count distribution, $f(S_{\rm measured})$, a framework where we have demonstrated KDE’s advantages in stability and fidelity. A full treatment to recover the true underlying distribution, $f(S_{\rm true})$, from the convolved data would require a formal statistical deconvolution. Applying advanced techniques, such as deconvolution KDE\citep[e.g.,][]{Yi2021}, to this problem is a valuable and necessary next step for future work, building upon the baseline KDE framework validated here.

\section{Conclusions}

We summarize the main findings of this work as follows.

\begin{enumerate}
    \item We developed a KDE-based framework for estimating differential radio source counts, including both fixed and adaptive bandwidth formulations with boundary correction. Compared to the conventional binned method, KDE provides a continuous, nonparametric estimate that avoids binning artifacts and performs better near survey flux limits. All computations in this study were performed using \texttt{AstroKDE}, a Python package we developed for astronomical KDE applications.

    \item Through controlled simulations based on a known luminosity function, we demonstrated that KDE methods, particularly the adaptive variant, consistently yield more accurate source count estimates than both fixed-width (Scott’s rule) and advanced (Bayesian Blocks) binned approaches, especially in the high-flux regime where data are sparse and uncertainty is high.

    \item We applied our framework to real observational data from the LoTSS Deep Fields \citep{2021A&A...648A...5M}. Our analysis, supported by both adaptive KDE and the Bayesian Blocks method, robustly confirms the main sub-mJy pronounced “drop and bump" feature, but demonstrates that a secondary, modest bump at ~ $\sim 10$ mJy in the Boötes field is likely an artifact of the original empirical binning choice.

    \item We highlighted the flexibility of KDE in accommodating observational incompleteness through weighted estimation, which applies weights continuously at the level of individual sources rather than averaging them in discrete bins. This makes it particularly well suited for modern and future radio surveys characterized by complex selection effects and large data volumes. Although this work focuses on radio source counts, the KDE-based framework is fully general and can be applied to source count analyses at other wavelengths, such as optical, infrared, or X-ray, where similar challenges arise.

\end{enumerate}

\begin{acknowledgments}
We thank the anonymous reviewer for the many constructive comments and suggestions, leading to a clearer description of these results. We acknowledge the financial support from Science Fund for Distinguished Young Scholars of Hunan Province (grant No. 2024JJ2040) and the National Natural Science Foundation of China (grant No. 12073069). Z.Y. is supported by the Xiaoxiang Scholars Programme of Hunan Normal University.
\end{acknowledgments}



%
%

\bibliography{ref}{}
\bibliographystyle{aasjournal}

\end{document}